\documentclass[aps,preprint,prd,epsfig,nofootinbib]{revtex4}
\pdfoutput=1
\usepackage{amssymb}
\usepackage{amsmath}
\usepackage{color}

\usepackage{fancyhdr}
\usepackage{slashed}
\usepackage[utf8]{inputenc}
\usepackage{graphicx}
\usepackage{epsfig}
\usepackage{amssymb}
\usepackage{bm}
\usepackage{color}
\usepackage[dvipsnames]{xcolor}
\newcommand{\be}{\begin{equation}}
\newcommand{\ee}{\end{equation}}
\newcommand{\bea}{\begin{eqnarray}}
\newcommand{\eea}{\end{eqnarray}}
\newcommand{\bes}{\begin{subequations}}
\newcommand{\ees}{\end{subequations}}
\newcommand{\nn}{\nonumber\\}
\newcommand{\bc}{\begin{center}}
\newcommand{\ec}{\end{center}}
\usepackage{amsmath, amssymb} 
\usepackage{slashed}
\usepackage{graphicx}
\usepackage{subcaption}
\begin{document}

\title{Naturally low scale type I seesaw mechanism and its viability in the 3-3-1 model with right-handed neutrinos }

\author{E. Cabrera$^{a}$, D. Cogollo$^{b,c}$, C. A. de S. Pires$^{d}$}
\email{cpires@fisica.ufpb.br}

\affiliation{{$^a$ Departamento de Física, Pontifícia Universidade Católica do Rio de Janeiro, 22452-970, Rio de Janeiro, RJ, Brazil},\\
{$^b$ Departamento de
F\'{\i}sica, Universidade Federal de Campina Grande, Caixa Postal 10071, 58109-970,
Campina Grande, PB, Brazil}, \\
{$^c$Northwestern University,  Department of Physics  and Astronomy,  2145  Sheridan Road,  Evanston,  IL 60208,  USA},\\
{$^d$  Departamento de Física, Universidade Federal da Paraíba, Caixa Postal 5008, 58051-970, Jo\~ao Pessoa, PB, Brazil}}

\date{\today}

\begin{abstract}
Seesaw mechanisms are the simplest and the  most elegant  way of generating small masses for the active neutrinos $(m_\nu)$. In these mechanisms  $m_\nu$ is inversely proportional to the lepton number breaking scale $(M)$ that, in the particular case of the type I seesaw mechanism,  is the Majorana mass of the right-handed neutrinos. In the canonical case right-handed neutrinos are supposed to be heavy  belonging to the GUT scale. With the advent of the LHC people began to suppose these neutrinos having mass at TeV scale. In this case very tiny Yukawa couplings are required. As far as we know there are no constraints on the energy scales  associated to the seesaw mechanisms. In what concern 3-3-1 models, when we trigger the type I seesaw mechanism   the lepton number breaking scale 
  that suppresses active neutrino masses contributes to the masses of the standard gauge bosons. Current data on $m_W$  demands the mechanism  to be performed at GeV scale. As main implication we may have right handed neutrinos with mass varying from few keVs up to hundreds of  GeVs. We also investigate the viability of the mechanism and found as interesting result that in the case in which the right-handed neutrino masses belong to the range keV-MeV scale,  viability of the mechanism demands that the lightest of the right-handed neutrinos be stable, which  makes of it a natural dark matter candidate, and that the lightest of the active neutrinos be much lighter than the other two active neutrinos.
\end{abstract}

\maketitle
\section{Introduction}

The canonical  expression for the neutrino masses provided by the type I seesaw mechanism,  $m_\nu \propto\frac{v^2}{M}$\cite{Minkowski:1977sc,Mohapatra:1979ia}, involves two energy parameters $v$ and $M$ with $M$ being a lepton-number breaking scale. In principle $v$ and $M$ are unknown parameters but the common assumption is that   $v\ll M$ which may leads to neutrino masses at eV scale  for certain values of $v$ and $M$. For example, for Yukawa coupling of order of unity  $M$ at GUT scale requires $v$ at the electroweak scale and $M$ at TeV scale requires $v$ at MeV.  This last case was promoted by  the advent of the LHC\cite{Kersten:2007vk}.   It would be great if we had a scenario that could shed some light on these energy parameters $v$ and $M$.

In its original version the $SU(3)_C \times SU(3)_L \times U(1)_N$(3-3-1) model with right-handed neutrinos(331RHN for short) involves three triplets of scalars\cite{Singer:1980sw,Foot:1994ym}. They  are not sufficient to generate correct mass for  the neutrinos. We can solve this problem by adding one  sextet of scalars to the particle content of the model which has the nice feature of generating  neutrino masses by means of the canonical  seesaw mechanisms\cite{Montero:1992jk,Dong:2008sw,Cogollo:2008zc,Ky:2005yq,Cogollo:2009yi,deSousaPires:2018fnl,Ferreira:2019qpf}.

In this work we show that, when the type I seesaw mechanism is implemented into the 331RHN by means of the sextet of scalars, the energy parameters $v$ and $M$ 
contribute to the masses of the standard  gauge bosons $W^{\pm}$ and $Z^0$. Current value of the mass of $W^{\pm}$  provides a constraint on the energy scale $M$ demanding that it  belongs to the GeV scale. As main implication of this we have that  right-handed neutrinos may develop masses in the range keV-GeV.  We  discuss  cosmological implications of  the lightest right-handed neutrinos and analyse its  contribution  to the neutrinoless double beta decay. 

This work is organized in the following way. In Sec. II we present the main aspects of the model and derive a constraint on the energy seesas scale. In Sec. III we develop the type I seesaw mechanism and discuss the profile of the mass and mixing of the neutrinos involved in the mechanism. In Sec. IV we present the implications of light right-handed neutrinos in the neutrinoless double beta decay and in Sec. V we present our conclusions.

\section{The main aspects of the  model  and the  constraint on the seesaw energy scale}

\subsection{Particle content and neutrino masses }
The leptonic content of the model is arranged in triplets and singlets of leptons  in the following form
\begin{equation}
f_{aL}= \begin{pmatrix}
\nu_{a}     \\
\ell_{a}       \\
\nu^{c}_{a} \\
\end{pmatrix}_{L} \sim (1,3,-1/3), \quad e_{aR}\sim (1,1,-1),
\end{equation}
with $a=e\,,\,\mu\,\,\tau$ representing the three SM generations of leptons.

In the hadronic sector, anomaly cancellation requires one family transforming differently from the other two\cite{Liu:1993gy,Ng:1992st}. This fact allows three possibles arrangements for the quark family\cite{Long:1999ij,Oliveira:2022vjo}. Here we present the arrangement where  the third generation comes in triplet and the other two come in an anti-triplet representation of $SU(3)_L$, 

\begin{eqnarray}
&&Q_{i_L} = \left (
\begin{array}{c}
d_{i} \\
-u_{i} \\
d^{\prime}_{i}
\end{array}
\right )_L\sim(3\,,\,\bar{3}\,,\,0)\,,u_{iR}\,\sim(3,1,2/3),\,\,\,\nonumber \\
&&\,\,d_{iR}\,\sim(3,1,-1/3)\,,\,\,\,\, d^{\prime}_{iR}\,\sim(3,1,-1/3),\nonumber \\
&&Q_{3L} = \left (
\begin{array}{c}
u_{3} \\
d_{3} \\
u^{\prime}_{3}
\end{array}
\right )_L\sim(3\,,\,3\,,\,1/3),u_{3R}\,\sim(3,1,2/3),\nonumber \\
&&\,\,d_{3R}\,\sim(3,1,-1/3)\,,\,u^{\prime}_{3R}\,\sim(3,1,2/3),
\label{quarks} 
\end{eqnarray}
where  the index $i=1,2$ is restricted to only two generations. The primed quarks are new heavy quarks with the usual $(+\frac{2}{3}, -\frac{1}{3})$ electric charges. 

The scalar sector of the 331RHN that lead to the type I seesaw mechanism is composed by three triplets and one sextet of scalars\cite{Ky:2005yq,Dong:2008sw,Cogollo:2008zc}:
\begin{eqnarray}
&&\eta = \left (
\begin{array}{c}
\eta^0 \\
\eta^- \\
\eta^{\prime 0}
\end{array}
\right ),\,\rho = \left (
\begin{array}{c}
\rho^+ \\
\rho^0 \\
\rho^{\prime +}
\end{array}
\right ),\,
\chi = \left (
\begin{array}{c}
\chi^0 \\
\chi^{-} \\
\chi^{\prime 0}
\end{array}
\right ), \nonumber \\
&&  S=\frac{1}{\sqrt{2}}\left(\begin{array}{ccc}
\sqrt{2}\, \Delta^{0} & \Delta^{-} & \Phi^{0} \\
\newline \\
\Delta^{-} & \sqrt{2}\, \Delta^{--} & \Phi^{-} \\
\newline \\
\Phi^{0} & \Phi^{-} & \sqrt{2}\, \sigma^{0} \end{array}\right)\,\,\,.
\label{scalarcont} 
\end{eqnarray}
with $\eta$ and $\chi$ transforming as $(1\,,\,3\,,\,-1/3)$, $\rho$ as $(1\,,\,3\,,\,2/3)$ and  $S$ transforming as $(1,6,-2/3)$. The sextet $S$ generates mass exclusively for the neutrinos. When only $\Delta^0$ and $\sigma^0$ develop VEVs neutrinos gain mass by means of the type II seesaw mechanism\cite{Cogollo:2009yi}. When only $\Phi^0$ and $\sigma^0$  develop VEVs neutrinos gain mass by means of the type I seesaw mechanism. We focus on the type I seesaw mechansim.

Before present the type I mechanism, let us present the Yukawa sector of the model which involves  the following terms:
\begin{eqnarray}
&-&{\cal L}^Y =f_{ij} \bar Q_{i_L}\chi^* d^{\prime}_{j_R} +f_{33} \bar Q_{3_L}\chi u^{\prime}_{3_R} + g_{ia}\bar Q_{i_L}\eta^* d_{a_R} +h_{3a} \bar Q_{3_L}\eta u_{a_R}\nonumber \\
&& +g_{3a}\bar Q_{3_L}\rho d_{a_R}+h_{ia}\bar Q_{i_L}\rho^* u_{a_R}+ G^{l}\bar f_{l_L} \rho e_{l_R} + G_{ab}^{s}(\overline{f_{L}^a})(f_{L}^b)^{c}S +  \mbox{H.c.},
\label{yukawa}
\end{eqnarray}
where $a=1,2,3$.  The discrete symmetry $\eta\,\,,\,\,\rho \rightarrow -(\eta\,\,,\,\,\rho)$ is sufficient to guarantee such simple Lagrangian. For the sake of simplicity, we considered the charged leptons in a diagonal basis. After spontaneous breaking of the symmetry by means of adequate choice of the VEVs such content of scalar generates masses for all massive particles of the model including fermions and gauge bosons. For our proposal here we assume that only  $\eta^0\,\,, \rho^0$, $\chi^{\prime 0}$, $\Phi^0$ and $\sigma^0$ develop VEV,
\begin{equation}
  \eta^0\,\,, \rho^0\,\,, \chi^{\prime 0}\,\,, \Phi^0\,\,, \sigma^0 =\frac{1}{\sqrt{2}}\Big(v_{\eta\,, \rho\,,\chi^{\prime}\,, \Phi^0\,, \sigma^0}+R_{_{\eta\,, \rho\,,\chi^{\prime}\,,\phi^0\,,\sigma^0}}+iI_{_{\eta\,, \rho\,,\chi^{\prime}\,, \phi^0\, ,\sigma^0 }}\Big)\,.
\end{equation}

This set of VEVs generates the type I seesaw mechanism.

As it was showed in Refs. \cite{Ky:2005yq,Cogollo:2008zc,Dong:2008sw} when $\Phi_{0}$ and $\sigma_{0}$ develop VEV the Yukawa interaction $G^{\nu}_{ab} \overline{f_{aL}}\, S\,(f_{bL})^c$ generates the type I seesaw mechanism for neutrino masses once neutrinos gain masses by means of the mass matrix
\begin{eqnarray}
\dfrac{1}{2}
\begin{pmatrix}
\overline{ \nu_{L}} & \overline{(\nu_{R})^{c}}
\end{pmatrix}
\begin{pmatrix}
0 & m_{D}^{T} \\
m_{D} & M_{R}
\end{pmatrix}
\begin{pmatrix}
(\nu_{L})^{c} \\
\nu_{R}
\end{pmatrix}+ \mbox{H.c.} =\frac{1}{2}\bar S_{L} M_\nu S_{L}^ C+ \mbox{H.c.},
\end{eqnarray}
where $S_{L}=(\nu_{L} , (\nu_{R})^{c})^{T}$ with $\nu_L=(\nu_{e_L}\,,\,\nu_{\mu_L} \,,\, \nu_{\tau_L})^T$ and $(\nu_{R})^{c}=((\nu_{e_R})^{c}\,,\,(\nu_{\mu_R})^{c}\,,\, (\nu_{\tau_R})^{c})^T$. 

As an interesting aspect, note that both $m_D$ and $M_R$ in $M_\nu$ have the same texture

\begin{eqnarray}
\label{diracmajorana}
m_{D}=v_{\phi}(G_{ab}), & M_{R}=\sqrt{2}v_{\sigma}(G_{ab}),
\end{eqnarray}
once both matrices share the same Yukawa coupling $G$. In other words, the mass of the heavy neutrinos are closely connected to the masses of the active neutrinos. This is a remarkable aspect of the model.

On assuming that $M_R \gg m_D$ the block diagonalization of this matrix leads to
\begin{equation}
m_{light}=-m_{D}^{T}M_{R}^{-1}m_{D}= -\frac{v^2_\Phi}{\sqrt{2} v_\sigma}G\,\,\,\,\,\,\,\,\mbox{and}\,\,\,\,\,\, M_{heavy}=\sqrt{2}v_\sigma G. 
\end{equation}
This is the canonical type I seesaw mechanism for generation of small masses for the active neutrinos. In it the VEV $v_\sigma$ is the lepton number breaking scale. There is no  bounds on this scale in the literature. We show below that this is not  the case of the type I seesaw mechanism generated in the 331RHN. 
\subsection{Gauge sector and the upper bound on $v_\sigma$}
The gauge sector of the 331RHN is composed by four charged gauge bosons, $W^{\pm}$ and $V^{\pm}$, four neutral gauge bosons, $U^0$, $U^{\dagger 0}$, $Z$, $Z^{\prime}$ and the photon $A$.  Their masses and mixing are obtained from the Lagrangian
\begin{equation}
    {\cal L}=\sum_\phi^{\eta,\rho,\chi} ({\cal D}_\mu \phi) ({\cal D}^\mu \phi)^{\dagger} + Tr[({\cal D}_\mu S) ({\cal D}^\mu S)^{\dagger} ]
    \label{heavyNH}
\end{equation}

As it is well developed in the literature, after  $\eta^0$, $\rho^0$ and $\chi^{\prime 0}$ develop VEVs,  the charged gauge bosons $W^{\pm}$ and $V^{\pm}$ gain the following mass expression $m^2_W=\frac{g^2}{4}(v_{\eta}^{2}+v_{\rho}^{2})$ and $m^2_{V}=\frac{g^2}{4}(v_{\chi^{\prime}}^{2}+v_{\rho}^{2})$\cite{Long:1995ctv}. The novelty here is that when $\Phi^0$ and $\sigma^0$ develop VEVs, they switch on a mixing between $W^{\pm}$ and $V^{\pm}$ given by

\begin{equation}
\label{misturaWV}
\dfrac{g^2}{4}
\begin{pmatrix}
W_{\mu} & V_{\mu}
\end{pmatrix}
\begin{pmatrix}
(v_{\eta}^{2}+v_{\rho}^{2}+v_{\Phi}^{2}) & \sqrt{2}v_{\Phi}v_{\sigma}\\
\sqrt{2}v_{\Phi}v_{\sigma} & (v_{\chi^{\prime}}^{2}+v_{\rho}^{2}+v_{\Phi}^{2}+2v_{\sigma}^{2})
\end{pmatrix}
\begin{pmatrix}
W^{\mu}\\
V^{\mu}
\end{pmatrix}.
\end{equation}

After diagonalizing this mass matrix we obtain

\begin{eqnarray}
    && m^2_{\tilde W} \approx \frac{g^2}{4}( v^2_\eta+v^2_\rho+v^2_\Phi+\frac{v^4_\sigma}{v^2_{\chi^{\prime}}}),\nonumber \\
    && m^2_{\tilde V} \approx \frac{g^2}{4}( v^2_{\chi^{\prime}}+v^2_\rho+v^2_\Phi+2v^2_\sigma).
    \label{CGBmasses}
\end{eqnarray}

The transformation we used to diagonalize  the mass matrix \eqref{misturaWV} is:

\begin{equation}
\label{wvtransformation}
\begin{pmatrix}
W_{\mu}\\
V_{\mu}
\end{pmatrix}
=
\begin{pmatrix}
C_{\theta} & S_{\theta}\\
-S_{\theta} & C_{\theta}
\end{pmatrix}
\begin{pmatrix}
\Tilde{W_{\mu}}\\
\Tilde{V_{\mu}}
\end{pmatrix},
\end{equation}

where the mixing angle  is given by:

\begin{equation}
    \tan 2\theta \sim \dfrac{2\sqrt{2}v_{\Phi}v_{\sigma}}{v_{\chi^{\prime}}^{2}}.
\label{tan2theta}
\end{equation}

Now comes the interesting point. Remember that $v^2_\eta +v^2_\rho \approx (246)^2$ GeV$^2$. Consequently, from the new contribution terms  for the mass of $\tilde W$ presented above we must have that  $v_\Phi \ll v_\eta\,,\, v_\rho$ and $v_\sigma < v_{\chi^{\prime}}$. For we have a more precise idea of the relation among $v_\sigma$ and $v_ {\chi^{\prime}}$, we resort to the recent CDF collaboration report that claims a new result for the mass of  $W^{\pm}$,  $m_{W_{CDF}}=80.4335 \pm 0.0094$ GeV\cite{CDF:2022hxs}, which deviated from the standard prediction $m_{W_{SM}}=80.357 \pm 0.006$ GeV\cite{deBlas:2022hdk}, at 7$\sigma$. Considering the central value of the prediction, the deviation is  $\Delta W=0.09635$ GeV. For $v_\Phi$ smaller than the other VEV's, as required by seesaw mechanism,  in such a way that its contribution to $m_W$ may be neglected, the term $\frac{g}{2}\frac{v^2_\sigma}{v_{\chi^{\prime}}}$ is solely responsible by the deviation. In this case, once the current LHC bound on $m_{Z^{\prime}}$ requires $v_{\chi^{\prime}}\approx 12$ TeV\cite{Coutinho:2013lta} or less\cite{Alves:2022hcp}, we then have that such deviation may be accommodated by the model for $v_\sigma \approx 60$ GeV at most. What is interesting here is that, once $v_\Phi$ and $v_\sigma$ are involved in the seesaw mechanism for neutrino masses, as discussed in the next section, such constraint on $v_\sigma$ is demanding that the seesaw mechanism be performed at GeV scale. This is nice because the signature of the  mechanism, the right-handed neutrinos, may be probed by many via as, for example, collider, astrophysical and cosmological data.

Moreover, for $v_\sigma \approx 60$ GeV, neutrino mass at eV scale requires $v_\Phi$ around $10^ {-5}$ GeV. In addition, as LHC constraint on the mass of $Z^{\prime}$ requires $v_{\chi^{\prime}} \approx 10^4$ GeV\cite{Coutinho:2013lta} implies $\tan 2 \theta \approx 10^{-11}$, which is very small and then may be neglected without any loss for the phenomenology of the model. From now on we  consider such  scenario.

For sake of completeness, we briefly discuss the impact of such VEV's in the gauge neutral sector composed by  $Z^0$, $Z^{\prime}$ and $U^0$. It is well known that  $Z^0$ and $Z^{\prime}$  mix between themselves giving rise to $Z_1$ and $Z_2$, see  Ref. \cite{Long:1995ctv}. The novelty here is that the VEVs $v_\Phi$ and $v_\sigma$ provide a mixing among $Z^0$, $Z^{\prime}$ and $U^0$ once the physical gauge bosons are obtained after diagonalize the mass matrix

\begin{equation}
\label{mistuneutros}
\dfrac{1}{2}
\begin{pmatrix}
Z_{\mu}^{0} & Z_{\mu}^{\prime} & U_{\mu}^{0}
\end{pmatrix}
\begin{pmatrix}
m_{Z^{0}}^2 & \delta_{Z^{0} Z^{\prime}}^2 & \delta_{Z^{0} U^{0\dagger}}^2\\
\delta_{Z^{\prime} Z^{0}}^2 & m_{Z^{\prime}}^2 & \delta_{Z^{\prime} U^{0\dagger}}^2\\
\delta_{U^{0} Z^{0}}^2 & \delta_{U^{0} Z^{\prime}}^2 & m_{U^{0}}^2
\end{pmatrix}
\begin{pmatrix}
Z^{0 \mu}\\
Z^{\prime\mu}\\
U^{0 \mu \dagger}
\end{pmatrix},
\end{equation}
with
\begin{eqnarray}
m_{Z^0}^2 & = & \frac{g^2}{4C_{W}^2}(v_{\eta}^{2}+v_{\rho}^{2}+v_{\Phi}^{2}),\nonumber\\
m_{Z^{\prime}}^{2} & = & \dfrac{g^{2}}{4h_{W}}[\dfrac{(v_{\rho}^2+v_{\Phi)}^2}{C_{W}^2}+\dfrac{v_{\eta}^{2}(1-2S_{W}^2)^2}{C_{W}^2}+4C_{W}^2(v_{\chi^{\prime}}^{2}+4v_{\sigma}^{2})],\nonumber\\
m_{U^{0}}^{2} & = & \dfrac{g^{2}}{2}(v_{\chi^{\prime}}^{2}+2v_{\sigma}^{2}+v_{\eta}^{2}+4v_{\Phi}^{2}),\nonumber\\
\delta_{Z^{0} Z^{\prime}}^{2} & = & \dfrac{g^{2}}{4C_{W}^2\sqrt{h_{W}}}[v_{\eta}^{2}(1-2S_{W}^2)-v_{\rho}^2-v_{\Phi}^2]=\delta_{Z^{\prime} Z^{0}}^{2},\nonumber\\
\delta_{Z^{0} U^{0 \dagger}}^{2} & = & \dfrac{g^{2}}{2C_{W}}(v_{\Phi}v_{\sigma})=\delta_{U^{0} Z^{0}}^{2},\nonumber\\
\delta_{Z^{\prime} U^{0 \dagger}}^{2} & = & -\dfrac{g^{2}}{2C_{W}\sqrt{h_{W}}}v_{\Phi}v_{\sigma}(1+4C_{W}^2)=\delta_{U^{0} Z^{\prime}}^{2}
\end{eqnarray}
where $h_{W}=3-4S_{W}^2$. 
Such mixing will have deep implications in flavor physics.

Concerning neutrino masses, the problem we must take care is with the masses of the heavy neutrinos determined by the expression $M_{heavy}=\sqrt{2}Gv_\sigma$. The Yukawa couplings $G$´s are determined by the mixing angles and squared mass differences of the standard neutrinos provided by atmospherics, solar and reactor neutrino oscillation experiments. Depending on the choice of $v_\Phi$ the values  of the Yukawa couplings  $G$´s may render heavy neutrinos with mass at GeV, MeV or keV scales. Heavy neutrinos with mass at GeV scale face no problem at all, but heavy neutrinos with mass at MeV scale must have lifetime bigger than the lifetime of the universe. If this is the case, such neutrino is a nice candidate for the dark matter of the universe. If this is not the case, such neutrino must have a lifetime around 0.1s in order to obey big bang nucleosynthesis (BBN) constraints. We check now under what conditions such cases can be realized.

\section{The type I seesaw mechanism and the profile of neutrino mass and mixing }\label{secIII}

Let us focus on the diagonalization of $M_\nu$ which is a $6 \times 6 $ mass matrix
with $G$ being a symmetric matrix formed by the Yukawa couplings $G_{ab}$. As it is very well known, the diagonalization of $M_\nu$, for the case $v_{\sigma} \gg v_{\Phi}$, leads to a relation among these VEVs known as the see-saw mechanism \cite{Mohapatra:1979ia,Minkowski:1977sc}. Such diagonalizaton is done in two stage. Firstly  $M_\nu$ is  block diagonalized. For this purpose let us define the matrix $W$,

\begin{equation}
\label{mixW}
W\simeq  
\begin{pmatrix}
1-\dfrac{1}{2}RR^{\dagger} & R\\
-R^{\dagger} & 1-\dfrac{1}{2}R^{\dagger}R
\end{pmatrix},
\end{equation}

where $R=m_{D}^{\dagger}(M_{R}^{\dagger})^{-1}$. Such matrix $W$ diagonalize $M_\nu$:

\begin{equation}
\label{blockdiagonal}
\dfrac{1}{2}W^{T}M_\nu W=\dfrac{1}{2}
\begin{pmatrix}
 m_{light} & 0 \\
 0 & M_{heavy}
\end{pmatrix},
\end{equation}
with $m_{light}=-m_{D}^{T}M_{R}^{-1}m_{D}= -\frac{v^2_\Phi}{\sqrt{2} v_\sigma}G$ and $M_{heavy}=M_{R}=\sqrt{2}v_\sigma G$. The second stage is the diagonalization of the block diagonalized matrix in Eq. \eqref{blockdiagonal} through the rotation matrix $U$,
\begin{equation}
\dfrac{1}{2}U^{T}
\begin{pmatrix}
 m_{light} & 0 \\
 0 & M_{heavy}
\end{pmatrix}
U=
\dfrac{1}{2}
\begin{pmatrix}
 m_{\nu} & 0 \\
 0 & M_{\nu},
\end{pmatrix}
\end{equation}
where, now, $m_\nu=\mbox{diag}(m_1\,,\,m_2\,,\,m_3)$ and $M_{\nu}=\mbox{diag}(M_1\,,\,M_2\,,\,M_3)$. Perceive that the job can be done with a unique unitary matrix $V=WU$, such that $V^{T}M_{\nu_{6 \times 6}}V=m_{diagonal_{6 \times 6}}$ with $U$ defined as,

\be 
U=\left(\begin{array}{cc}
U_{PMNS} & 0\\ 
 0 &  U_R
      \end{array}\right),
      \label{matrixfinal2}
\ee
with $U_{PMNS}$ being  the well kown PNMS matrix that diagonalizes $m_{light}$ while $U_R$  diagonalizes $M_{heavy}$. In the end of the day we have that  $M_\nu$ is directly diagonalized by the following rotation matrix $V$ given by,

\begin{equation}
\label{mixV}
V\simeq  
\begin{pmatrix}
\left(1-\dfrac{1}{2}RR^{\dagger}\right)U_{PMNS} & RU_R\\
-R^{\dagger}U_{PMNS} & \left(1-\dfrac{1}{2}R^{\dagger}R \right)U_R
\end{pmatrix}.
\end{equation}

For simplicity, we will define the matrix $V$ by the following  form:

\begin{equation}
V=\begin{pmatrix} { V }^{ \nu \nu  } & { V }^{ \nu N } \\ { V }^{ N\nu  } & { V }^{ NN } \end{pmatrix}.
\label{Vmatrix}
\end{equation}

The rotation matrix $V$ relates  the flavor basis  $S_{L}=(\nu_{L} , (\nu_{R})^{c})^{T}$ with the  physical basis $\hat S_{L}=(n_1,n_2,n_3,N_1,N_2,N_3)^{T}$ given by $S_{L}=V\hat S_{L}$. 

\begin{equation}
\label{flavormass}
\begin{pmatrix}
\nu_{L} \\
(\nu_{R})^{c}
\end{pmatrix}
=
\begin{pmatrix} { V }^{ \nu \nu  } & { V }^{ \nu N } \\ { V }^{ N\nu  } & { V }^{ NN } \end{pmatrix}
\begin{pmatrix}
n_{L} \\
(N_{R})^{c}
\end{pmatrix}
\end{equation} 

 We now discuss the viability of such mechanism. Our worry here is with the mass of the heavy neutrinos which obey the expression $M_{heavy}=\sqrt{2}Gv_\sigma$. For $v_\sigma$ at tens of GeVs the masses of the heavy neutrinos depend of the Yukawa coupling $G$ which will be determined the physics of the active neutrinos. That is all about thetory of type I seesaw mechanism. From now on we insert data.

Neutrino oscillation experiments provide us with three mixing angles and two mass squared differences. Neglecting CP-phases, according to current data the central values of the mixing angles are\cite{ParticleDataGroup:2022pth} 
\begin{equation}
    \theta_{12} \approx 35^o,\,\,\,\, \theta_{23} \approx 45^o\,\,\,\,\, \mbox{and}\,\,\,\,\, \theta_{13} \approx 8.5^o .
    \label{angles}
\end{equation}
This translates in the following pattern of neutrino mixing  $U_{PMNS}$
\begin{eqnarray}
U_{PMNS}=\left(\begin{array}{ccc}
0.81 & 0.57 & 0.15 \\
\newline \\
-0.49 & 0.52 & 0.69 \\
\newline \\
0.32 & -0.64 & 0.69 \end{array}\right).
\label{UPMNS}
\end{eqnarray}

Unlike neutrino mixing where the three angles are very well determined, neutrino mass experiments cannot determine the absolute values of the three neutrino masses, they provide two quadratic mass squared $|\Delta m_{\mbox{sol}}^2| \approx(7.4 - 7.9)\times 10^{-5}\mbox{eV}^2$ and $|\Delta m_{\mbox{atm}}^2|\approx2.5\times 10^{-3}\mbox{eV}^2$ \cite{ParticleDataGroup:2022pth} which allow us fix only two neutrino masses. This fact generates an enormous amount of arbitrariness in neutrino physics.  In our case we have seven free parameters, namely six Yukawa couplings, $G$´s, and one VEV $v_\Phi$(We are assuming that $v_\sigma$ is fixed by $m_{W}$ discussed above), and five constraints, namely three angles and two quadratic mass squared.  Then once we have more free parameters than constraints, in order to continue we have to consider scenarios. It is this approach we are going to apply next.

For the neutrino masses we use as illustrative example the following set of eigenvalues:

\begin{eqnarray}
   &&  m_1 \approx 3 \times 10^{-4}\mbox{eV},\,\,\, m_2 \approx 8.6 \times 10 ^{-3}\mbox{eV}\,\,\,\mbox{and}\,\,m_3\approx 0.05\mbox{eV},
\label{eigenvalues}
\end{eqnarray}
which provides $|\Delta m_{\mbox{sol}}^2| \approx(7.4 - 7.9)\times 10^{-5}\mbox{eV}^2$ and $|\Delta m_{\mbox{atm}}^2|\approx2.5\times 10^{-3}\mbox{eV}^2$.

Next, we will analyze three different mass regimes for the right handed neutrinos and discuss the physical implications of  each case.

\subsection{GeV Scale RHN}
\label{GevScale}
According to previous section, $v_\sigma$ must lie at $\sim 60$ GeV scale if we want that our model solves the recently reported W boson mass anomaly\cite{CDF:2022hxs}. Moreover $v_\sigma$ is the energy scale of lepton number violation, in other words, it is the energy scale that suppresses neutrino masses in the type I seesaw mechanism. The masses of the heavy neutrinos (RHN) are given by  $M_{heavy}=\sqrt{2}v_\sigma G$, then, it is natural to have RHN at the GeV scale. 

On the other hand, as showed above, light neutrino masses take the simple form $m_\nu \approx -\frac{v^2_\Phi}{\sqrt{2}v_\sigma}G$. By assuming that $\frac{v^2_\Phi}{\sqrt{2}v_\sigma} $ is of the order of $\sim 1.5\times10^{-11}$ GeV, it will imply in  $v_\Phi \approx 3.5\times10^{-5}$ GeV. For such scenario recovers neutrino mass and mixing showed above we need, for example, the following set of Yukawa couplings:

\begin{eqnarray}
G \approx-\left(\begin{array}{ccc}
0.270051 & 0.510939 & 0.149391 \\
\newline \\
0.510939 & 1.83167 & 1.4805 \\
\newline \\
0.149391 & 1.4805 & 1.9088 \end{array}\right),
\label{Gevcouplings}
\end{eqnarray}

This scenario accommodates standard nutrino physics data. Let us see the implications on the RHNs. The set of Yukawa couplings in \eqref{Gevcouplings} implies for  $M_{heavy}=\sqrt{2}v_\sigma G$, 

\begin{eqnarray}
M_{heavy}\approx\left(
\begin{array}{ccc}
 -22.9146 & -43.3546 & -12.6762 \\
 -43.3546 & -155.422 & -125.625 \\
 -12.6762 & -125.625 & -161.967
\end{array}
\right)\mbox{GeV}
\label{MRNH}
\end{eqnarray}

whose eigenvalues are:
\begin{equation}
    |M_1|\approx 1.69 \; \mbox{GeV},\,\,\,\,\, |M_2|\approx48.64 \; \mbox{GeV}\,\,\,\,\mbox{and}\,\,\,|M_3|\approx282.843 \;\mbox{GeV}.
    \label{MR-NH}
\end{equation}

Note that this illustrative case posses right-handed neutrinos with masses varying from  few GeV´s up to some hundred of GeV´s. The mixing among the right-handed neutrinos and the active ones is given by,

\begin{eqnarray}
    V^{\nu N}\approx
    \left(
\begin{array}{ccc}
 3.40 & 2.39 & 0.63 \\
 -2.06 & 2.18 & 2.90 \\
 1.34 & -2.69 & 2.90
\end{array}
\right)\times10^{-7}.
\label{VvNGevscale}
\end{eqnarray}

In view of all this we must check if our scenario respect current constraints. The first constraint  we consider relates to the values of the matrix $\frac{1}{2}RR^{\dagger}$ which measures the deviation from unitarity of the $U_{PNMS}$ matrix\cite{Fernandez-Martinez:2016lgt}. In our case the order of magnitude of all the elements of the matrix $\frac{1}{2}RR^{\dagger}$ is about $\sim 10^{-14}$ satisfying from far the requirement. Also, the neutrinoless double beta decay imposes constraints on the matrix \eqref{VvNGevscale}. For neutrinos in the GeV scale $\frac{\sum_{i}U_{ei}^{2}}{M_{i}} \leq 10^{-8}$\cite{Benes:2005hn}. In our illustrative examples this constraint  is also  fulfilled from far. As for the rare decay $\mu \longrightarrow e\gamma$ exist the constraint $Br(\mu \longrightarrow e\gamma) \leq 4.2\times 10^{-13}$\cite{MEG:2013oxv}. In our model $\mbox{Br}(\mu \longrightarrow e\gamma) \sim 10^{-31}$.

In summary, the scenario  where the heavy neutrinos belong to the GeV scale with mass varying from few GeVs up to hundredes of GeVs is completely viable and these heavy neutrinos may be probed at the LHC by means of the processes $pp \rightarrow Z^{\prime} \rightarrow N N \rightarrow l^{\pm}l^{\mp}l^{\mp}\nu_{ljj}$ or  $pp \rightarrow Z^{\prime} \rightarrow N N \rightarrow l^{\pm}l^{\pm}W^{\mp}W^{\mp}$\cite{Freitas:2018vnt,Das:2017deo,Das:2017flq,Cox:2017eme}. All this means that such scenario is  promissing and interesting.

\subsection{MeV and keV right-handed neutrino}
\label{MevKevscale}

Here we discuss scenarios where the lightest of the heavy neutrinos, $N_1$, has mass at keV or MeV scale. This is quite possible  since $v_\sigma$  lies  at GeV 
scale or lower. Nevertheless, in what follow we keep assuming $v_\sigma =60$ GeV. We then show that  our seesaw model is able to generate RHN in both scales and that in both cases  a DM candidate is possible if the active neutrino $m_1$ is very light, which is quite plausible from the experimental point of view. 

Let us first consider that our lightest RHN ($N_1$) has mass inside the range $(1.023 \lesssim M_1 \leq 100)$ MeV\footnote{For $N_1$ above 100 MeV it decays very fast recovering practically the GeV case discussed above.}. In this case $N_1$ decays through the channels $N_{1} \longrightarrow \nu_{\alpha}e^{-}e^{+}$ (with contributions of both neutral and charged currents, with $\alpha=e^{-},\mu^{-}, \tau^{-}$), $N_{1} \longrightarrow \nu\nu\bar{\nu}$ (only through the neutral current), and $N_{1} \longrightarrow \nu_{\alpha}\gamma$ ($\alpha=e^{-},\mu^{-}, \tau^{-}$). The first two widths are the dominant ones. The first width is derived from the general expression\cite{Gorbunov:2007ak,Berryman:2017twh}:

\begin{eqnarray}
\Gamma \left(N_{1} \to \nu_\alpha \ell_\beta^- \ell^+_\beta \right) & = &  \frac{G_F^2 M_1^5 |V^{\nu N}_{\alpha 1}|^2}{192\pi^3} \cdot \Big[ \Big(C_1 (1-\delta_{\alpha\beta}) + C_3 \delta_{\alpha\beta} \Big) \times  \nonumber \\
& & \left. \Big( \left(1- 14y_\ell^2 - 2 y_\ell^4  - 12 y_\ell^6 \right) \sqrt{1-4 y_\ell^2}+ 12 y_\ell^4 (y_\ell^4 -1) L \Big)  \right. \nonumber \\
& & + 4 \Big(C_2 (1-\delta_{\alpha\beta}) + C_4 \delta_{\alpha\beta} \Big) \times \Big(y_\ell^2 \left(2+10y_\ell^2-12y_\ell^4 \right) \sqrt{1-4 y_\ell^2} \nonumber \\
& & + 6 y_\ell^4 \left(1 - 2y_\ell^2 + 2 y_\ell^4 \right) L \Big) \Big], 
\label{gamma1}
\end{eqnarray}

with, $\left( y_\ell \equiv \frac{m_{\ell_\beta}}{M_{1}} \right)$, and:

\begin{eqnarray}
C_1 = \frac{1}{4} \left(1-4 \sin^2 \theta_w + 8 \sin^4 \theta_w \right), & & C_2 = \frac{1}{2} \sin^2 \theta_w \left(2 \sin^2 \theta_w - 1 \right), \nonumber \\
C_3 = \frac{1}{4} \left(1+4 \sin^2 \theta_w + 8 \sin^4 \theta_w \right), & & C_4 = \frac{1}{2} \sin^2 \theta_w \left(2 \sin^2 \theta_w + 1 \right), \nn
\label{functions}
\end{eqnarray}
where $\theta_w$ is the Weinberg angle, and
\begin{equation}
L = \log \left[ \frac{1-3y_\ell^2 - \left(1- y_\ell^2 \right) \sqrt{1-4y_\ell^2}}{y_\ell^2 \left(1+\sqrt{1-4y_\ell^2} \right)} \right].
\label{logaritmo}
\end{equation}

As for the second and third width, we have:

\begin{equation}
\Gamma \left(N_{1} \to \sum_{\beta}\nu_{\alpha}\nu_{\beta}\bar{\nu_{\beta}}\right) = \frac{G_F^2 M_1^5 |V^{\nu N}_{\alpha 1}|^2}{192\pi^3}\,\ ,
\label{gammaN2}
\end{equation}

and 

\begin{eqnarray}
\Gamma \left(N_{1} \to \nu_\alpha \gamma \right) & = & \frac{9  \alpha_{EM} G_F^2 M_{1}^{5}|V^{\nu N}_{\alpha 1}|^2}{512 \pi^2} ,
\label{gamma3}
\end{eqnarray}

respectively. The total width considered by us is:

\begin{equation}
\Gamma=2\times\left(\sum_{\alpha}\Gamma \left(N_{1} \to \sum_{\beta}\nu_{\alpha}\nu_{\beta}\bar{\nu_{\beta}}\right)+\sum_{\alpha}\left(\Gamma_ {\nu_{\alpha}e^{-}e^{+}}+\Gamma_{\nu_{\alpha}\gamma}\right)\right),
\label{total}
\end{equation}

We show below  that  bounds on $\Gamma$ translate into a relationship between $M_1$ and $m_1$. 

Let us first discuss the case when the mass of $N_1$ lies at MeV scale. In this case, for the model be safe, $N_1$ needs to have lifetime bigger than the universe ($\tau \gtrsim 10^{17}$ s) or decay fast with a lifetime  $\tau \lesssim 0.1$ s\cite{Ruchayskiy:2012si}. We check now if such cases are phenomenologicaly viables. The care we must have is due to the fact  that active and heavy neutrinos share the same Yukawa coupling $G$ which means that both mass matrices $m_{light}$ and $M_{heavy}$ are diagonalized by $U_{PMNS}$. In other words, $U_R=U_{PMNS}$. In order to be more precise, let us proceed in the following way: suppose the existence in the model of $N_{1}$ with mass at MeV scale and decaying with a total width \eqref{total} (this means that $(1.023 \lesssim M_1 \leq 100)$ MeV). In our model $V^{\nu N}_{\alpha 1}=\frac{v_\phi}{\sqrt{2}v_\sigma}(U_R)_{\alpha 1}$,  $m_\nu \approx -\frac{v^2_\Phi}{\sqrt{2}v_\sigma}G$ and $M_{heavy} \approx \sqrt{2}v_\sigma G$. Observe that the eigenvalue  $m_{1}$ can be written as $m_{1}=\frac{v^2_\Phi}{\sqrt{2}v_\sigma}(U_{R}^{T}GU_{R})_{11}$ and isolating $v_\Phi$ we get  that $v_{\phi}^{2}=\dfrac{m_{1}\sqrt{2}v_\sigma}{(U_{R}^{T}GU_{R})_{11}}$, which allow us to write
$(V^{\nu N}_{\alpha 1})^{2}=\dfrac{m_{1}(U_R)^{2}_{\alpha 1}}{\sqrt{2}v_\sigma(U_{R}^{T}GU_{R})_{11}}=\dfrac{m_{1}(U_R)^{2}_{\alpha 1}}{M_{1}}$. Replacing  $V^{\nu N}_{\alpha 1}$ in  the total width \eqref{total} above we have that the total width (or the lifetime $\tau=\frac{1}{\Gamma}$) depends exclusively on $M_1$ and $m_1$. We allow $M_1$ varying in the range $(1.023 \lesssim M_1 \leq 100)$MeV and ask if $m_1$  at sub-eV scale is compactible with  $N_1$  decaying before  big bang nucleosynthesis(BBN) or being stable.  We used  $G_F=1.166\times 10^{-5}$GeV$^{-2}$ and $\alpha_{EM}=\dfrac{e^2}{4\pi}$. We first  check for what values of $m_{1}$ are compactibles with $N_1$ decaying before big bang nucleosynthesis (BBN). As interesting results we obtain that  for $N_1$ (with $M_1$ in the range $(1.023 \lesssim M_1 \leq 100)$ MeV)  decaying before BBN it demands an active neutrino with mass above eV scale. For example, for $M_1 \approx 1.023$ MeV a lifetime of $\tau \lesssim 0.1$ s requires  $m_1 \gtrsim 126.65$ GeV, and at the other extreme, when $M_1 \approx 100$ MeV, a lifetime of $\tau \lesssim 0.1$s requires $m_1 \gtrsim 0.957$ keV.  Then the mechanism developed here does not support $M_1$ in the range $(1.023 \lesssim M_1 \leq 100)$ MeV respecting BBN. This hypothesis is discarded.

The other possibility is  $N_1$ stable. If this is the case, the model may have an interesting dark matter candidate with mass belonging to the MeV scale. We can check that this is possible for $m_1$ very light. For example, for an $N_{1}$ with $M_1\approx10$ MeV to be stable, it demands $m_1 \lesssim 9.6\times10^{-12}$ eV. Nevertheless, very tiny $m_1$ may require large Yukawa couplings, disfavoring the implementation of the seesaw mechanisms in the model. Of course that this depend of the value of $v_\sigma$. The case we are discussing here, with $v_\sigma =60$ GeV, the model favors a scenario with $N_1$ having mass around  1 MeV. To see this observe that, for $M_1\approx1.023$ MeV to be stable demands $m_1 \lesssim 1.26\times10^{-7}$ eV. This is quite plausible because neutrino oscillation experiments support at least one very light or even null neutrino eigenvalue. This is an interesting result because we are constraining the viability of  the type I seesaw to the existence of dark matter in the form of RHN with mass belonging to the MeV scale (more precisely among ($1.023 \lesssim M_1 \leq 100)$MeV), and of this being associated to a very light active neutrino.

Let us present an illustrative scenario for $M_{1} \approx 1.023$ MeV and $m_1 \approx 1.26\times10^{-7}$ eV. First, for $m_2$ and $m_3$ we used the same set of values in \eqref{eigenvalues}, as well as the same mixing matrix \eqref{UPMNS} and mixing angles \eqref{angles} and $v_{\sigma}=60$ GeV. Second, the relationship  $\frac{m_1}{M_1}=\dfrac{v_{\Phi}^2}{(\sqrt{2}v_{\sigma})^{2}}$ implies $v_{\Phi} \approx 2.9\times10^{-5}$ GeV. Taking all this into account, the Yukawa couplings are:

\begin{eqnarray}
G\approx-\left(\begin{array}{ccc}
0.368799 & 0.744819 & 0.207011 \\
\newline \\
0.744819 & 2.62189 & 2.12943 \\
\newline \\
0.207011 & 2.12943 & 2.7367 \end{array}\right),
\label{Mevcouplings}
\end{eqnarray}

This set of Yukawa couplings  implies 

\begin{eqnarray}
M_{heavy}\approx
\left(
\begin{array}{ccc}
 -31.2937 & -63.2 & -17.5655 \\
 -63.2 & -222.475 & -180.688 \\
 -17.5655 & -180.688 & -232.217 \\
\end{array}
\right)
\mbox{GeV}
\label{MRNHMev}
\end{eqnarray}

whose eigenvalues are:

\begin{equation}
    |M_1|\approx 0.001023 \; \mbox{GeV},\,\,\,\,\, |M_2|\approx69.8238 \; \mbox{GeV}\,\,\,\,\mbox{and}\,\,\,|M_3|\approx405.952 \;\mbox{GeV}.
    \label{MR-NHMev}
\end{equation}

In this case, the active-RHN  mixing is:

\begin{eqnarray}
    V^{\nu N}\approx
\left(
\begin{array}{ccc}
2.84271 & 2.00043 & 0.526428 \\
-1.71966 & 1.82495 & 2.42157 \\
1.12305 & -2.24609 & 2.42157 \\
\end{array}
\right)
\times10^{-7},
\label{VvNMevscale}
\end{eqnarray}

The set of eigenvalues in \eqref{MR-NHMev} and the mixing matrix in \eqref{VvNMevscale} guarantee that $N_1$ is in agreement with current constraint. For example, such scenario predicts  $\mbox{Br}(\mu \longrightarrow e\gamma) \sim 10^{-31}$ which is far from current bound.

Now, let us deal with the case for $(100 \leq M_1 \leq 1000)$ keV. In this case, $N_{1}$ decays only through the channels $N_{1} \longrightarrow \nu\nu\bar{\nu}$, and $N_{1} \longrightarrow \nu_{\alpha}\gamma$ ($\alpha=e^{-},\mu^{-}, \tau^{-}$). As in the previous case, for $N_{1}$  decaying before BBN, it demands an active neutrino with mass above eV scale. For $M_{1} \approx 100$ keV, for example, a lifetime of $\tau \lesssim 0.1$ s requires a value of $m_1 \gtrsim 1.3\times 10^6$ GeV, and for $M_{1} \approx 1000$ keV, a lifetime of $\tau \lesssim 0.1$ s requires a value of $m_1 \gtrsim 133$ GeV, then, besides the seesaw mechanism is able to generate  $N_1$ with mass in the range $(100 \leq M_1 \leq 1000)$ keV in the model, such possibility does not respect the BBN bounds. On the other hand, stable candidates are possibles. For example, for $N_1$  with $M_1 \approx 150$K keV to be stable demands $m_1 \lesssim 2.64\times10^{-4}$ eV. As in the previous case, we present an illustrative scenario for $M_{1} \approx 150$ keV and $m_{1} \approx 2.64\times10^{-4}$ eV. For $m_2$ and $m_3$ we used again the same set of values in \eqref{eigenvalues}, as well as the same mixing matrix \eqref{UPMNS} and mixing angles \eqref{angles} and $v_{\sigma}=60$ GeV. The Yukawa couplings we found are:

\begin{eqnarray}
G\approx-\left(\begin{array}{ccc}
0.0000269489 & 0.0000514267 & 0.0000149359 \\
\newline \\
0.0000514267 & 0.000183909 & 0.000148746 \\
\newline \\
0.0000149359 & 0.000148746 & 0.000191695 \end{array}\right),
\label{Kevcouplings}
\end{eqnarray}

which implies a $v_{\Phi} \approx 3.5\times10^{-3}$ GeV, and a mass matrix $M_{heavy}$:

\begin{eqnarray}
M_{heavy}\approx
\left(
\begin{array}{ccc}
 -0.00228669 & -0.0043637 & -0.00126735 \\
 -0.0043637 & -0.0156052 & -0.0126215 \\
 -0.00126735 & -0.0126215 & -0.0162659 \\
\end{array}
\right)
\mbox{GeV}
\label{MRNHKev}
\end{eqnarray}

whose eigenvalues are:

\begin{equation}
    |M_1|\approx 150 \; \mbox{KeV},\,\,\,\,\, |M_2|\approx4.88 \; \mbox{MeV}\,\,\,\,\mbox{and}\,\,\,|M_3|\approx28.40 \;\mbox{MeV}.
    \label{MR-NHKev}
\end{equation}

As for the mixing matrix $V^{\nu N}$ we obtain:

\begin{eqnarray}
    V^{\nu N}\approx
\left(
\begin{array}{ccc}
 3.39814 & 2.39128 & 0.62928 \\
 -2.05567 & 2.18152 & 2.89471 \\
 1.34248 & -2.68495 & 2.89471 \\
\end{array}
\right)
\times10^{-5},
\label{VvNKevscale}
\end{eqnarray}

In summary, the cases of RHN with masses in the range of keV or MeV are physically viable only for when the lightest of the RHN has a lifetime bigger than the age of the universe. In other words, the viability of the scenarios demand the existence of a dark matter candidate which is a nice result. Moreover, new interactions (self-interactions) among active neutrinos existent in the 331RHN may allow non-thermal production of $N_1$  by means of the Dodelson-Widrow mechanism to constitute 100$\%$ of the dark matter of the universe\cite{DeGouvea:2019wpf}. In the case of $N_1$ belonging to MeV scale its production follow the procedure done in Ref.\cite{Dutra:2021lto}.

\section{Neutrinoless double beta decay implications}

In this section we analyse the contributions to the neutrinoless double beta decay due to the presence of the RHN that we studied in the previous sections. We follow the procedure developed in \cite{Schechter:1981bd}. As is well know, RHN mixing with active neutrinos contribute to this rare processes and the magnitude of the contribution depends on the size of the mixing and also on the mass of the RHN. Moreover, besides the standard $W$ boson charged current, the model has a novel boson $V$ that also mediates charged currents. In order to clarify the idea, let us show the total charged current from which we derived all the contributions to the process\cite{Dias:2005jm},

\begin{eqnarray}
\ensuremath{\mathcal{L}}^{331_{RH\nu}}_{CC} & = &
\dfrac{-g}{\sqrt{2}}\{\bar{\ell}_{aL}\gamma^{\mu}\nu_{aL}W_{\mu}^{-} +\bar{\ell}_{aL}\gamma^{\mu}(\nu_{aR})^{c}V_{\mu}^{-}+\bar{d}_{aL}\gamma^{\mu}u_{aL}W_{\mu}^{-}+\mbox{H.c.}\},
\label{correntecarregada331}
\end{eqnarray}
with $a=1,2,3$. Before proceeding, it is important to remember that $W$ and $V$ mix among themselves. The transformation \eqref{wvtransformation} was used to diagonalize  the mass matrix \eqref{misturaWV}. It happens that the prediction $\tan 2\theta\sim 10^{-11}$ is very small,  which allow us to neglect any contribution to the process whose amplitude is proportional to $\tan\theta$ or $\sin \theta$. Considering this, let us  firstly write \eqref{correntecarregada331} in terms of the physical eigenstates of the gauge bosons
\begin{eqnarray}
\ensuremath{\mathcal{L}}^{331_{RH\nu}}_{CC} & = &
\dfrac{-g}{\sqrt{2}}\{\bar{\ell}_{aL}\gamma^{\mu}\nu_{aL}C_{\theta}\Tilde{W_{\mu}}^{-}+\bar{\ell}_{aL}\gamma^{\mu}\nu_{aL}S_{\theta}\Tilde{V_{\mu}}^{-}\nonumber\\ 
 & & +\bar{\ell}_{aL}\gamma^{\mu}(\nu_{aR})^{c}(-S_{\theta})\Tilde{W_{\mu}}^{-}+\bar{\ell}_{aL}\gamma^{\mu}(\nu_{aR})^{c}C_{\theta}\Tilde{V_{\mu}}^{-}
 \nonumber\\
  & & + \bar{d}_{aL}\gamma^{\mu}u_{aL}C_{\theta}\Tilde{W_{\mu}}^{-}+\bar{d}_{aL}\gamma^{\mu}u_{aL}S_{\theta}\Tilde{V_{\mu}}^{-}+\mbox{H.c.}
 \}.
\label{correntecarregada331b}
\end{eqnarray}

The first and third line of \eqref{correntecarregada331b} are contributions coming from the currents involving the $W$ boson, and we will call these currents of $J^{\mu}$. The second line is a current generated by the $V$ boson and we will call this current of $(J^{\mu})^{c}$. Applying \eqref{flavormass} on \eqref{correntecarregada331b} we obtain an expression of the charged current in the physical base. For leptons we have:

\begin{eqnarray}
J^{\mu}_{\ell} & = &
\dfrac{-g}{\sqrt{2}}\{\bar{\ell}_{aL}\gamma^{\mu}V^{\nu\nu}_{ak}n_{kL}C_{\theta}\Tilde{W_{\mu}}^{-}+\bar{\ell}_{aL}\gamma^{\mu}V^{\nu N}_{am}(N_{mR})^{c}C_{\theta}\Tilde{W_{\mu}}^{-}
\nonumber\\ 
 & & +\bar{\ell}_{aL}\gamma^{\mu}V^{\nu\nu}_{ak}n_{KL}S_{\theta}\Tilde{V_{\mu}}^{-}+\bar{\ell}_{aL}\gamma^{\mu}V^{\nu N}_{am}(N_{mR})^{c}S_{\theta}\Tilde{V_{\mu}}^{-}
\},
\label{jleptons}
\end{eqnarray}

\begin{eqnarray}
(J^{\mu})^{c}_{\ell} & = &
\dfrac{-g}{\sqrt{2}}\{\bar{\ell}_{aL}\gamma^{\mu}V^{N \nu}_{ak}n_{kL}(-S_{\theta})\Tilde{W_{\mu}}^{-}+\bar{\ell}_{aL}\gamma^{\mu}V^{N N}_{am}(N_{mR})^{c}(-S_{\theta})\Tilde{W_{\mu}}^{-}
\nonumber\\ 
 & & +\bar{\ell}_{aL}\gamma^{\mu}V^{N \nu}_{ak}n_{kL}C_{\theta}\Tilde{V_{\mu}}^{-}+\bar{\ell}_{aL}\gamma^{\mu}V^{N N}_{am}(N_{mR})^{c}C_{\theta}\Tilde{V_{\mu}}^{-}
\},
\label{jleptonsc}
\end{eqnarray}

and for the charged currents involving the quarks, we obtained:

\begin{eqnarray}
J^{\mu}_{q} & = &
\dfrac{-g}{\sqrt{2}}\{\bar{d}_{aL}\gamma^{\mu}u_{aL}C_{\theta}\Tilde{W_{\mu}}^{-}+\bar{d}_{aL}\gamma^{\mu}u_{aL}S_{\theta}\Tilde{V_{\mu}}^{-}
\}.
\label{jquarks}
\end{eqnarray}

With all these currents at hand we proceed to calculate the contributions to the neutrinoless double beta decay in each one of the three scenarios we discussed in section \eqref{secIII}. The standard contribution involves the exchange of light neutrinos through the $J^{\mu}$ current mediated by the $\Tilde{W_{\mu}}^{-}$ boson (this means the first term of equation \eqref{jleptons}) \footnote{The light neutrinos we are referring  here are the active neutrinos}, and its amplitude is:

\begin{equation}
\label{SC}
\ensuremath{\mathcal{A}}_{(\beta\beta)0\nu} (1) = 8G_{F}^{2}C_{\theta}^{4}\frac{\langle m_{\nu}\rangle}{q^{2}}   
\end{equation}

with

\begin{equation}
\label{mnu}
\langle m_{\nu}\rangle = \left\vert \sum_{i=1}^{3} (V^{\nu\nu}_{ei})^{2}m_{i} \right\vert,
\end{equation}
and $q^{2}$ being the momentum transfer which is of the order of $\sim$ (100-200) MeV for the $0\nu\beta\beta$ process.

The second dominant contribution is similar to standard one, but considering the exchange of heavy neutrinos through the $J^{\mu}$ current mediated by $\Tilde{W_{\mu}}^{-}$ boson (this means the second term of equation \eqref{jleptons})\footnote{The heavy neutrinos we are referring to here are the RHN with masses from keV up to GeV}. Its amplitude must be divided in two parts, depending on $q \gg M$ or $q \ll M$. For the case $q \ll M$ the amplitude is given by:

\begin{equation}
\label{C2}
\ensuremath{\mathcal{A}}_{(\beta\beta)0\nu} (2) = -8G_{F}^{2}C_{\theta}^{4} \left\langle \frac{1}{M_{\nu}}\right\rangle ,
\end{equation}

where

\begin{equation}
\label{Mnu}
\left\langle \frac{1}{M_{\nu}}\right\rangle = \left\vert\sum_{i=1}^{3}\frac{(V_{ei}^{\nu N})^{2}}{M_{i}}\right\vert,
\end{equation}

and for the case $q \gg M$ the amplitude is:

\begin{equation}
\label{c3}
\ensuremath{\mathcal{A^{\prime}}}_{(\beta\beta)0\nu} (2) = 8G_{F}^{2}C_{\theta}^{4}\frac{\langle M_{\nu}\rangle}{q^{2}}   
\end{equation}

with

\begin{equation}
\label{Mnu2}
\langle M_{\nu}\rangle = \left\vert \sum_{i=1}^{3} (V_{ei}^{\nu N})^{2}M_{i} \right\vert.
\end{equation}

All the the other contributions to the process depend on $\sin \theta$ or $\tan \theta$, and as we showed, are negligible. For the sake of completeness we show all then in appendix \ref{A}, but we did not analyse them. 

Now, realize that $\dfrac{\ensuremath{\mathcal{A}}_{(\beta\beta)0\nu} (2)}{\ensuremath{\mathcal{A}}_{(\beta\beta)0\nu} (1)}$ will be always a number multiplied by $q^{2}$. This analysis can be done for the case when the RHN are all in the GeV scale. Then, for the set of parameters studied in the section \eqref{GevScale} the ratio between the amplitudes is:

\begin{equation}
    \frac{\ensuremath{\mathcal{A}}_{(\beta\beta)0\nu} (2)}{\ensuremath{\mathcal{A}}_{(\beta\beta)0\nu} (1)} = - 1.68\times 10^{-2} \mbox{GeV}^{-2}q^2.
\label{amplitudes1}
    \end{equation}
For  $q \sim 200$Mev we get  $ \frac{\ensuremath{\mathcal{A}}_{(\beta\beta)0\nu} (2)}{\ensuremath{\mathcal{A}}_{(\beta\beta)0\nu} (1)} \sim  10^{-4}$. As expected, RHN with mass at GeV scale gives a tiny contribution to the neutrinoless double beta decay.

For the MeV case studied in the section \eqref{MevKevscale}, there are two different amplitudes that we must take into account. For the lightest RHN ($M_{1} \sim 1.023$ MeV)  the amplitude is of the form  given in \eqref{c3}, and when compared with the standard one we found that it is $\sim 10^{5}$ times smaller. The second amplitude is due to the presence of the other two RHN whose eigenvalues are $M_{2} \sim 69.82$ GeV and $M_{3} \sim 405.95$ GeV. In this case the amplitude is of the form given in  \eqref{C2} and for an estimated value of $q \sim 200$ MeV, the amplitude is $\sim 10^{6}$ times smaller than the standard one, too.

Finally, for the keV case studied  in the section \eqref{MevKevscale}, the amplitude, that is of the form given in \eqref{c3} (since all the eigenvalues \eqref{MR-NHKev} are $\ll q$),  may be of the same order of magnitude that the standard one. For our illustraive example we got

\begin{equation}
    \frac{\ensuremath{\mathcal{A^{\prime}}}_{(\beta\beta)0\nu} (2)}{\ensuremath{\mathcal{A}}_{(\beta\beta)0\nu} (1)} \sim 1.00064,
\label{amplitudes1}
    \end{equation}
which is an interesting result. All this contributes to the enrichment of the 331RHN model.

\section{Conclusions}

In this work we revisited the implementation of  the type I seesaw mechanism into the 331RHN. As new element we observed that the mechanism must be performed at low energy scale, more precisely around tens of GeVs but can be lower.  As far as we know this result is a novelty in the literature.

When the sextet of scalars is added to the 331RHN, the type I seesaw mechanism can be triggered with specific choices of the VEVs of the neutral components of the sextet. It happens that the VEV that generates Majorana mass for the right-handed neutrinos (the one that is responsible by the suppression of the active neutrino masses in the type I seesaw mechanism) also contributes to the mass of the gauge bosons of the model, and, in particular, to the mass of the standard charged gauge boson $W^{\pm}$. Consequently such VEV much lies at GeV scale, more precisely around tens of GeV if we wish to explain the recent CDF result concerning $W$ mass. 

As particular feature of the 331RHN  the right-handed neutrinos as well as left-handed ones compose the same leptonic triplet and the neutrino masses is achieved by means of the Yukawa interaction $G \bar f^C S^* f$. As result, the mass expression for active neutrinos as well as for the right-handed neutrinos share the same Yukawa couplings $G$. This fact connects the masses of the active neutrinos to the masses of the right-handed ones and has deep implications in the viability of the mechanism. Once the mechanism is performed at GeV scale, then right-handed neutrinos may develop masses in the range from few  keVs up to GeV scale. In the case of GeV scale, there is no problem at all and the mechanism is perfectly viable. In the case when at least one right-handed neutrino (the lightest of them)  belongs to keV or MeV scale we showed that the mechanism is viable only if the lightest right-handed neutrino is stable. This is very interesting because stability is a conditions to a neutral particle   be a  dark matter candidate of the universe. Another peculiar fact is that such stability require that the lightest of the active neutrino be very light. Then this scenario  has two signature, namely  the existence of  right-handed neutrino as dark matter and a very light active neutrino. Future experiments seeking for absolute neutrino mass and the search for the discovery of the nature of dark matter may probe this proposal. We also checked the contribution of the right-handed neutrinos to the neutrinoless double beta decay, which may receive a robust contribution from RHN with masses at keV scale. In summary, the type I seesaw mechanism when implemented into the 331RHN model has remarkable implications in physics beyond the standard model.

\appendix

\section{More contributions to the Neutrinoless double beta decay}
\label{A}

The following sketch was made for the case where RHN fulfill the condition $q \ll M$, and we called them heavy neutrinos, while active neutrinos are called light neutrinos. For the case when $q \gg M$, a subdivision of the amplitudes must be done as we did in the text of the article. Said that, we proceed as follow:

There is one  contribution involving $\Tilde{W_{\mu}}^{-}$ boson, but now interacting with the two charged currents $J^{\mu}$ and $(J^{\mu})^C$ exchanging  light neutrinos :

\begin{equation}
    \label{C3}
    \ensuremath{\mathcal{A}}_{(\beta\beta)0\nu} (3) = 8G_{F}^{2} \frac{\langle m_{\nu}^{\prime} \rangle}{q^{2}}C_{\theta}^{3}S_{\theta} 
\end{equation}

with 

\begin{equation}
    \langle m_{\nu}^{\prime} \rangle = \sum_{i=1}^{3} V^{\nu\nu}_{ei} V^{N\nu}_{ei}m_{i}
\end{equation}

Similar to the previous contribution, we have another new one but now exchanging  heavy neutrinos 

\begin{equation}
    \label{C4}
    \ensuremath{\mathcal{A}}_{(\beta\beta)0\nu} (4) = -8G_{F}^{2} C_{\theta}^{3}S_{\theta} \left\langle \frac{1}{M_{\nu}^{\prime}} \right\rangle
\end{equation}

with

\begin{equation}
    \left\langle \frac{1}{M_{\nu}^{\prime}} \right\rangle = \sum_{i=1}^{3} \frac{V_{ei}^{\nu N} V_{ei}^{NN}}{M_{i}}
\end{equation}

Let us consider contributions  with $\Tilde{W}$ via $J^{\mu}$  and $\Tilde{V}$ via $J^{\mu C}$ . In this case, for light neutrinos exchangedwe have:

\begin{equation}
    \label{C5}
    \ensuremath{\mathcal{A}}_{(\beta\beta)0\nu} (5) = 8G_{F}^{2}\left(\frac{m_{\Tilde{W}}}{m_{\Tilde{V}}}\right)^{2} \frac{\langle m_{\nu}^{\prime}\rangle}{q^{2}} C_{\theta}^{3}S_{\theta}  ,
\end{equation}

and with heavy neutrinos exchanged:

\begin{equation}
    \label{C6}
    \ensuremath{\mathcal{A}}_{(\beta\beta)0\nu} (6) = -8G_{F}^{2}\left(\frac{m_{\Tilde{W}}}{m_{\Tilde{V}}}\right)^{2} \left\langle \frac{1}{M_{\nu}^{\prime}} \right\rangle C_{\theta}^{3}S_{\theta}  
\end{equation}

As the last two contributions we show one involving only  $\Tilde{V}$  but exchanging  light neutrinos

\begin{equation}
    \label{C7}
    \ensuremath{\mathcal{A}}_{(\beta\beta)0\nu} (7) = 8G_{F}^{2}\left(\frac{m_{\Tilde{W}}}{m_{\Tilde{V}}}\right)^{4} \frac{\langle m_{\nu}^{\prime\prime} \rangle}{q^{2}} S_{\theta}^{2}C_{\theta}^{2}  
\end{equation}

where 

\begin{equation}
    m_{\nu}^{\prime\prime} = \sum_{i=1}^{3} (V_{ei}^{N\nu})^{2}m_{i}
\end{equation}

and the other exchanging heavy neutrinos :

\begin{equation}
    \label{C8}
    \ensuremath{\mathcal{A}}_{(\beta\beta)0\nu} (8) = -8G_{F}^{2}\left(\frac{m_{\Tilde{W}}}{m_{\Tilde{V}}}\right)^{4} \left\langle \frac{1}{M_{\nu}^{\prime\prime}} \right\rangle S_{\theta}^{2}C_{\theta}^{2}  
\end{equation}

with 

\begin{equation}
    \left\langle \frac{1}{M_{\nu}^{\prime\prime}} \right\rangle = \sum_{i=1}^{3} \frac{(V_{ei}^{NN})^{2}}{M_{i}}
\end{equation}

\section*{Acknowledgments}
C.A.S.P  was supported by the CNPq research grants No. 311936/2021-0. E.C. thanks the support received by the CNPq scholarship No. 140121/2022-6. D.C thanks the Department of Physics and Astronomy at Northwestern University for the hospitality during this period in which the project was carried out, especially André de Gouvêa for the useful discussions, time and motivation to develop this work. 
\bibliography{bibliography}
\end{document}